\documentclass{optica-article}
\articletype{Research Article}

\journal{opticajournal} 

\newcommand\ee{\mathrm{e}}
\newcommand\ii{\mathrm{i}}
\newcommand\cc{\mathrm{c}}
\newcommand\TT{\mathrm{T}}

\newcommand\dd{\,\mathrm{d}}
\newcommand\nm{\,\mathrm{nm}}
\newcommand\mm{\,\mathrm{mm}}
\newcommand\um{\,\mathrm{\mu m}}
\newcommand\ps{\,\mathrm{ps}}
\newcommand\fs{\,\mathrm{fs}}

\newcommand\sinv{\,\mathrm{s}^{-1}}
\newcommand\sgn{\operatorname{sgn}}

\begin{document}

\title{On-chip spatiotemporal optical vortex generation using an integrated metal-dielectric resonator}

\author{
Artem I. Kashapov\authormark{1,2}, 
Leonid L. Doskolovich\authormark{1,2,*}, 
Evgeni A. Bezus\authormark{1,2}, 
Dmitry A. Bykov\authormark{1,2}, and
Victor A. Soifer\authormark{1,2}
}

\address{\authormark{1}Image Processing Systems Institute -- Branch of the Federal Scientific Research Centre ``Crystallography and Photonics'' of Russian Academy of Sciences, 151 Molodogvardeyskaya st., Samara 443001, Russia\\}
\address{\authormark{2}Samara National Research University, 34 Moskovskoe shosse, Samara 443086, Russia\\}

\email{\authormark{*}leonid@ipsiras.ru}

\begin{abstract}
We theoretically demonstrate the possibility of generating a spatiotemporal optical vortex (STOV) beam in a dielectric slab waveguide.
The STOV is generated upon reflection of a spatiotemporal optical pulse from an integrated metal-dielectric structure consisting of metal strips ``buried'' in the waveguide.
For describing the interaction of the incident pulse with the integrated structure, we derive its ``vectorial'' spatiotemporal transfer function (TF) describing the transformation of the electromagnetic field components of the incident pulse.
We show that if the TF of the structure corresponds to the TF of a spatiotemporal differentiator with a $\pi/2$ phase difference between the terms describing temporal and spatial differentiation, then the envelope of the reflected pulse will contain an STOV in all nonzero components of the electromagnetic field.
The obtained theoretical results are in good agreement with the results of rigorous numerical simulation of the STOV generation using a three-strip metal-dielectric integrated structure.
We believe that the presented results pave the way for the research and application of STOV beams in the on-chip geometry.
\end{abstract}


\section{Introduction}\label{sec:intro}
An optical vortex (OV) beam is a special type of monochromatic optical beams possessing a zero in the field amplitude and a screw-type phase dislocation.
The most common type of OV beams corresponds to a beam with a phase singularity in the plane perpendicular to the propagation direction.
Such a beam possesses the so-called longitudinal orbital angular momentum (OAM) and is characterized by a rotational flow of energy density around the phase singularity.
OV beams with longitudinal OAM and the methods of their generation using various diffractive structures and metasurfaces have been extensively studied~\cite{1,2,3,4,5,6,7,8,9} and have many important applications including optical trapping~\cite{5,6}, super-resolution microscopy~\cite{7}, and free-space telecommunications~\cite{8,9}, among others.

Recently, a new class of OV beams has been discovered in optics, referred to as spatiotemporal optical vortex (STOV) beams~\cite{10,11,12,13,14,15,16,17,18,19,20,21}.
In contrast to the conventional OV beams, STOV beams are essentially polychromatic and carry orbital angular momentum, which is orthogonal to the beam propagation direction.
The presence of a transverse OAM attracts considerable research attention to the STOV beams due to their potential uses in optical trapping, optical communications and information processing.
Despite the very recent progress in the generation of STOV beams~\cite{11,12,13,14,15,18,19}, many issues related to the generation and investigation of such beams have not yet been completely studied.
In particular, an important feature of the STOV beams is that, unlike the conventional OV beams, they can exist in 2D geometry (the corresponding 2D space containing a transverse spatial coordinate and time or two spatial coordinates, one of which corresponds to the propagation direction)~\cite{21}.
This provides an intriguing new possibility for the generation and application of STOV beams on different integrated photonic platforms.
In particular, in a wide class of planar (integrated) optoelectronic systems, optical information processing is performed in a slab waveguide~\cite{22,23,24,25,26,27,28}.
Such geometry corresponds to the ``insulator-on-insulator'' platform and is suitable for the creation of fully integrated optical devices, in which the data carriers are the optical pulses and beams propagating in a slab waveguide layer.

In the present work, we for the first time theoretically investigate the generation of an STOV beam in a dielectric slab waveguide.
In the proposed approach, the STOV beam is generated upon reflection of a spatiotemporal optical pulse with a Gaussian envelope from an integrated metal-dielectric structure consisting of metal strips ``buried'' in the waveguide layer.
The interaction of the incident pulse with the metal-dielectric structure is described in the framework of the theory of linear systems.
We obtain a ``vectorial'' spatiotemporal transfer function (TF) describing the transformation of the components of the electromagnetic field of the incident pulse that occurs upon reflection from a metal-dielectric structure.
We demonstrate that if the TF of the structure corresponds to the TF of a spatiotemporal differentiator (i.\,e., has a linear form with respect to the spatial and angular frequencies), and the complex coefficients at the linear terms of the TF have a $\pi/2$ phase (argument) difference, then the envelope of the reflected pulse will contain an STOV in all nonzero components of the electromagnetic field.
The obtained theoretical results are confirmed by the results of the rigorous numerical simulation of the STOV generation in a slab waveguide using a three-strip metal-dielectric integrated structure and of its subsequent propagation.

\section{Representation of a spatiotemporal pulse}
Let us start by considering a general representation of a spatiotemporal optical pulse propagating in a dielectric slab waveguide.
Such a pulse can be represented as a superposition of slab waveguide modes with different angular frequencies and propagation directions.
Therefore, let us first write the dispersion relation of a mode propagating in a waveguide layer located at $|z| \leq a/2$, where $a$ is the waveguide thickness.
For the sake of simplicity, let us consider a symmetric waveguide with equal dielectric permittivities of the claddings.
We will focus on the fundamental transverse electric (TE) polarized mode of the waveguide, which is even.
Dispersion relation of even TE-polarized modes, from which their effective refractive indices ${n_{\rm eff}}$ can be found, has the form~\cite{29}
\begin{equation}
\label{eq:1}
k_{z,1} = \ii k_{z,2} \tan (k_{z,2} a / 2),
\end{equation}
where $\pm k_{z,l} = \pm (\omega/\cc) \sqrt {\varepsilon_l - n_{\rm eff}^2}$, $l = 1,2$ are the $z$-components of the wave vectors of the plane waves constituting the mode in the claddings and inside the waveguide layer, $\omega$ is the angular frequency of the mode, $\cc$ is the speed of light in vacuum, and $\varepsilon_1$ and $\varepsilon_2$ are the dielectric permittivities of the claddings and the waveguide layer, respectively.
The components of the electromagnetic field of a mode with the wave vector $(k_x, k_y) = \left(\sqrt{k_\parallel^2 - k_y^2}, k_y\right)$, where $k_\parallel = (\omega / \cc) n_{\rm eff}$ is the wavenumber of the mode, can be easily obtained from the Maxwell’s equations as
\begin{equation}
\label{eq:2}
{\bf\Phi}(x,y,z,t) = {\bf\Psi}(\omega, k_y, z){\ee^{\ii x\sqrt {k_\parallel^2 - k_y^2} + \ii k_y y}} \ee^{-\ii\omega t},
\end{equation}
where ${\bf\Phi} = \left[E_x \; E_y \; E_z \; H_x \; H_y \; H_z\right]^\TT$ is a column vector containing the mode field components.
Here, the vector function ${\bf\Psi}(\omega, k_y, z)$ represents the amplitudes of the field components and has the form
\begin{equation}
	\label{eq:3}
	{\bf\Psi}(\omega, k_y, z) = 
		\left\{ 
			\begin{aligned}
				&{\bf\Psi}_{\rm core}(\omega, k_y, z),\; &|z| < \frac{a}{2}, \\ 
     &{\bf\Psi}_{\rm cl}(\omega, k_y, z),\; &|z| > \frac{a}{2}, \\ 
			\end{aligned}
		\right.
\end{equation}
where the functions ${\bf\Psi}_{\rm core}$ and ${\bf\Psi}_{\rm cl}$ define the amplitudes of the field components inside the waveguide core layer and in the claddings, respectively:
\begin{equation}
	\label{eq:4}
	{\bf\Psi}_{\rm core}(\omega, k_y, z) = \cos (k_{z,2}z)
	\begin{bmatrix}
   -k_y (\omega/\cc) / k_\parallel^2 \\ 
    k_x (\omega/\cc) / k_\parallel^2 \\ 
    0 \\ 
   -\ii{k_x}{k_{z,2}}\tan ({k_{z,2}}z)/k_\parallel^2 \\ 
   -\ii{k_y}{k_{z,2}}\tan ({k_{z,2}}z)/k_\parallel^2 \\ 
   1 \\ 
	\end{bmatrix},
	\hspace{4.8em}
\end{equation}
\begin{equation}
	\label{eq:5}
{\bf\Psi}_{\rm cl}(\omega, k_y, z) = \cos (a{k_{z,2}}/2)
	\begin{bmatrix}
		-k_y (\omega/\cc) / k_\parallel^2 \\ 
		 k_x (\omega/\cc) / k_\parallel^2 \\ 
		 0 \\ 
		-\sgn (z) k_x k_{z,1}/k_\parallel^2 \\ 
		-\sgn (z) k_y k_{z,1}/k_\parallel^2 \\ 
		 1 \\ 
	 \end{bmatrix}
\ee^{\ii k_{z,1} (|z| - a/2)},
\end{equation}
where $\sgn(z)$ is the signum function.
Let us note that since for a waveguide mode, $n_{\rm eff}^2 > \varepsilon_1$, the quantity $k_{z,1}$ is purely imaginary, and the function ${\bf\Psi}_{\rm cl}(\omega, k_y, z)$ exponentially decays when moving away from the waveguide.
Let us also note that for a TE-polarized mode, $E_z \equiv 0$ and the corresponding third component of the vector ${\bf\Psi}(\omega, k_y, z)$ is equal to zero.
At $k_y = 0$, the mode propagates along the $x$ axis, and the field components $E_x$ and $H_y$ also vanish.

Next, let us consider a spatiotemporal optical pulse with a central frequency $\omega_0$ propagating in the waveguide along the $x$ axis.
Let us represent this pulse as a superposition of guided modes with different angular frequencies $\omega = \omega_0 + \tilde\omega$ and spatial frequencies $k_y$.
In this case, according to Eq.~\eqref{eq:2}, the envelope of the incident pulse is described by the following integral:
\begin{equation}
\label{eq:6}
\begin{aligned}
	{\bf P}(x,y,z,t) = \iint 
	&G(k_y,\tilde\omega){\bf \Psi }	\left( \omega_0+\tilde\omega,k_y, z \right)
	\\&\cdot\ee^{\ii x\left( k_x(\tilde\omega, k_y) - \frac{\omega_0}{\cc} n_{\rm eff}(\omega_0) \right) + \ii k_y y}
	\ee^{-\ii\tilde\omega t}
	\dd{k_y}\dd{\tilde\omega},
\end{aligned}
\end{equation}
where $G(k_y, \tilde\omega)$ is the pulse spectrum representing the amplitudes of the modes constituting the pulse and
\begin{equation}
\label{eq:7}
{k_x}(\tilde\omega, k_y) = \sqrt {{{\left( {\frac{\omega_0+\tilde\omega}{\cc}{n_{\rm eff}}(\omega_0+\tilde\omega)} \right)}^2} - k_y^2}
\end{equation}
are the $x$-components of the wave vectors of the modes.
Note that according to the dispersion relation of the mode~\eqref{eq:1}, the effective refractive index of the mode is considered as a function of frequency: $n_{\rm eff} = n_{\rm eff}(\omega_0+\tilde\omega)$.
The quantities $k_{z,l},\;l = 1,2$, and ${k_\parallel }$ in the expressions for the vector ${\bf \Psi}$~\eqref{eq:3}--\eqref{eq:5} depend on the angular frequency $\omega = \omega_0 + \tilde\omega$ as well.

By choosing the spectrum $G(k_y, \tilde\omega)$, one can provide a required form of one of the components of the pulse envelope at $x=0$ and at a certain fixed $z$.
In the following discussion, we will assume that the envelope of the $z$-component of the magnetic field $P_{H_z}$ is defined at $x=0$ in the central plane of the waveguide (at $z=0$).
Since ${\Psi}_{H_z}(\omega, k_y, 0) = 1$ [see Eq.~\eqref{eq:4}], we obtain:
\begin{equation}
\label{eq:8}
P_{H_z}(0,y,0,t) = \iint G(k_y, \tilde\omega ){\ee^{\ii k_y y}}{\ee^{ - \ii\tilde\omega t}}\dd{k_y}\dd\tilde\omega
\end{equation}
and, accordingly,
\begin{equation}
\label{eq:9}
G(k_y, \tilde\omega ) = \frac1{4\pi^2}\iint P_{H_z}(0,y,0,t) \ee^{-\ii k_y y} \ee^{\ii\tilde\omega t}\dd{y}\dd{t}.
\end{equation}

\section{Transformation of the pulse upon reflection and spatiotemporal optical differentiation}\label{sec:2}
\begin{figure*}[thb]
\centering\includegraphics{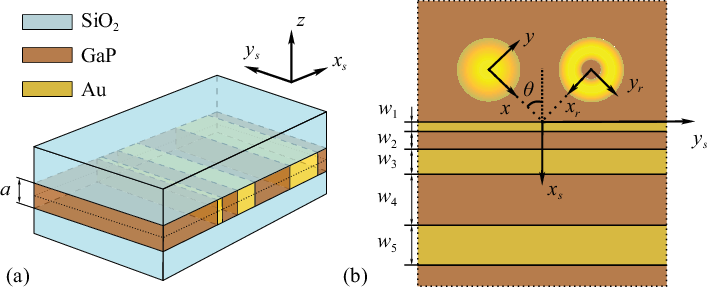}
\caption{\label{fig:1}Geometry of the considered integrated structure consisting of several metal strips ``buried'' in the core layer of a symmetric dielectric slab waveguide (left) and its central ($z=0$) cross-section (right). In the right panel, the generation of a spatiotemporal optical vortex is schematically shown.
}
\end{figure*}

Let us consider the transformation of the envelope of the incident pulse~\eqref{eq:6} occurring upon reflection of the pulse from a diffractive structure integrated into the waveguide (Fig.~\ref{fig:1}).
In what follows, as such a structure, we will consider a set of metal strips ``buried'' in the waveguide layer (Fig.~\ref{fig:1}).
We assume that the angle of incidence of the pulse $\theta$ is chosen to be sufficiently large so that for all the modes constituting the incident signal, the component of the wave vector, which is parallel to the metal strips (and is thus conserved upon reflection) is greater than the wavenumbers of both the plane waves in the claddings and the transverse magnetic (TM) polarized mode supported by the waveguide.
In this case, the energy of the incident TE polarized modes will be divided between the reflected and transmitted guided modes having the same polarization, a part of the energy also being absorbed due to the presence of the metal strips in the considered structure.
It is easy to show that this ``scattering suppression'' condition can always be fulfilled by choosing the angle of incidence provided that, as in the example considered in the present work, the fundamental TE-polarized guided mode is used as the incident wave~\cite{30,31}.
 
Upon reflection of the pulse from the structure, the amplitudes of the guided modes constituting the pulse are multiplied by the reflection coefficient $R(k_{y,s}, \omega_0+\tilde\omega)$, where $k_{y,s}$ is the tangential component of the wave vector of the incident mode written in the coordinate system $(x_s, y_s)$ associated with the structure [see Fig.~\ref{fig:1}(b)]:
\begin{equation}
\label{eq:10}
\begin{aligned}
k_{y,s} 
&= k_x\sin\theta + k_y\cos\theta 
\\&= \sin\theta\sqrt {{{\left( {\frac{\omega_0+\tilde\omega}{\cc}{n_{\rm eff}}(\omega_0+\tilde\omega)} \right)}^2} - k_y^2} + k_y\cos\theta.
\end{aligned}
\end{equation}
Similarly to the description of the incident pulse, which was performed in the coordinate system $(x, y)$ associated with it, it is convenient to represent the reflected pulse in its local coordinate system $(x_r, y_r)$ [Fig.~\ref{fig:1}(b)].
In this system, the reflected pulse is propagating in the negative direction of the $x_r$ axis and the modes constituting the pulse have the same $k_y$ components as in the incident pulse.
Thus, in the coordinate system $(x_r, y_r)$, the envelope of the reflected pulse has the form
\begin{equation}
\label{eq:11}
\begin{aligned}
{\bf P}_r(x_r, y_r, z, t) = 
\iint &G(k_y, \tilde\omega ){\bf\Psi}\left(\omega_0+\tilde\omega, k_y,z \right)R({k_{y,s}},\omega_0+\tilde\omega)
\\&\cdot\ee^{-\ii x_r\left( k_x(\tilde\omega, k_y) - \frac{\omega_0}{\cc}{n_{\rm eff}}(\omega_0) \right) + \ii k_y y_r}
\ee^{-\ii\tilde\omega t}
\dd{k_y}\dd\tilde\omega.
\end{aligned}
\end{equation}
From Eqs.~\eqref{eq:6} and~\eqref{eq:11}, it follows that the transformation of the envelope of the $H_z$ component of the incident pulse $P_{H_z}(0,y,0,t) \to P_{r, H_z}(0,y_r,0,t)$ corresponds to the transformation of the signal $P_{H_z}(0,y,0,t)$ by a linear system with the following transfer function:
\begin{equation}
\label{eq:12}
H(k_y, \tilde\omega ) = R(k_{y,s}(k_y, \tilde\omega), \omega_0+\tilde\omega).
\end{equation}

Let us assume that the used diffractive structure possesses a reflection zero at $k_y = 0$ and $\tilde\omega = 0$ ($\omega = \omega_0$).
We will also consider the case, in which the spectrum of the incident pulse $G(k_y, \omega)$ is ``concentrated'' in the vicinity of the point of zero reflection $(k_y, \tilde\omega ) = (0,0)$, so that in this vicinity the TF can be approximated by the linear expansion terms as
\begin{equation}
\label{eq:13}
H_{\rm lin}(k_y, \tilde\omega ) = c_y k_y + c_t \tilde\omega,
\end{equation}
where $c_y = \frac{\partial H}{\partial k_y}(0,0)$, $c_t = \frac{\partial H}{\partial \tilde\omega}(0,0)$.
Using the representation~\eqref{eq:13} and the properties of the Fourier transform, it is easy to obtain from Eqs.~\eqref{eq:6} and~\eqref{eq:11} that
\begin{equation}
\label{eq:14}
P_{r,H_z}(0,y,0,t) = \left( -\ii{c_y}\frac{\partial }{\partial y} + \ii{c_t}\frac{\partial }{\partial t} \right){P_{H_z}}(0,y,0,t).
\end{equation}
Therefore, a structure possessing a reflection zero optically implements the operation, which can be referred to as the spatiotemporal differentiation of the envelope of the component $H_z$ at $z=0$.
From Eqs.~\eqref{eq:11} and~\eqref{eq:13}, one can also obtain that in the used linear approximation, the envelope of the component $H_z$ at different $z$ values, as well as the envelopes of other components of the reflected pulse are proportional to the envelope of the component $H_z$ at $z=0$.
Indeed, by expanding the product ${\bf\Psi}_{\rm core}( \omega_0+\tilde\omega, k_y, z) H(k_y, \tilde\omega)$ from Eq.~\eqref{eq:11} up to the linear terms with respect to $k_y$ and $\tilde\omega$, we obtain
\begin{equation}
\label{eq:15}
\begin{aligned}
{{\bf{\Psi }}_{\rm core}}&\left( \omega_0+\tilde\omega, k_y, z \right)H(k_y, \tilde\omega ) 
\\&\approx \cos \left( k_{z,2}(\omega_0)z \right)\begin{bmatrix}
  0 \\ 
  1 / n_{\rm eff}(\omega_0) \\ 
  0 \\ 
  - \frac{\ii{k_{z,2}} \tan\left( k_{z,2}(\omega_0)z \right)}{n_{\rm eff}(\omega_0) \omega_0/\cc} \\ 
  0 \\ 
  1 \\ 
 \end{bmatrix}\left( c_y k_y + c_t\tilde\omega \right),
\end{aligned}
\end{equation}
where $k_{z,2}(\omega_0) = (\omega_0/\cc) \sqrt{\varepsilon_l - n_{\rm eff}^2(\omega_0)}$.
Thus, in the linear approximation~\eqref{eq:15}, the reflected pulse will have nonzero $E_y$ and $H_x$ components proportional to the envelope of the $H_z$ component at each fixed $z$.
As for the $H_z$ component itself, its values at nonzero values of $z$ are determined by the function $\cos \left(k_{z,2}(\omega_0)z \right)$.
Let us note that similar reasoning is true also for the field outside the waveguide core layer, which can be demonstrated by considering the linear expansion of the product ${\bf\Psi}_{\rm cl}\left( \omega_0 + \tilde\omega, k_y, z\right)H(k_y, \tilde\omega )$.

\section{Generation of a spatiotemporal optical vortex}
Let us theoretically show that a spatiotemporal optical vortex can be generated using a diffractive structure having a TF of a spatiotemporal optical differentiator~\eqref{eq:13}.
Following Ref.~\cite{20}, let us assume that the envelope of the $z$-component of the magnetic field of the incident pulse $P_{H_z}$ is described by a Gaussian function
\begin{equation}
\label{eq:16}
	P_{H_z}(y,0,t) = \exp \left\{ - \frac{y^2}{\sigma_y^2} 
	- \frac{t^2}{\sigma_t^2} \right\}.
\end{equation}
In this case, the spectrum of the envelope is also Gaussian:
\begin{equation}
\label{eq:17}
G(k_y, \tilde\omega ) = \frac{\sigma_y \sigma_t}{4\pi }\exp \left\{ - \frac{\sigma_y^2 k_y^2}{4} - \frac{\sigma_t^2{{\tilde\omega}^2}}{4} \right\}.	
\end{equation}
Let the TF of the used diffractive structure have the form of Eq.~\eqref{eq:13} with the coefficients $c_y$ and $c_t$ satisfying the relations
\begin{equation}
\label{eq:18}
\varphi = \arg (c_t/c_y) = \pm \pi/2,
\end{equation}
\begin{equation}
\label{eq:19}
	|c_t \sigma_y| \big/ |c_y \sigma_t| = 1.
\end{equation}
In this case, using Eqs.~\eqref{eq:14},~\eqref{eq:15},~\eqref{eq:4}, and~\eqref{eq:5}, we obtain
\begin{equation}
\label{eq:20}
P_{r, H_z}(0, \tilde y, z, \tilde t) = 2\ii\frac{c_y}{\sigma _y}
( \tilde y \mp \ii \tilde t)\ee^{ -{\tilde y}^2 - {\tilde t}^2}
F(z),
\end{equation}
where $\tilde y = {y_r}/{\sigma_y}$ and $\tilde t = t/{\sigma_t}$ are the normalized coordinates, and the function
\begin{equation}
\label{eq:21}
	F(z) = 
	\left\{ \begin{aligned}
  &\cos ({k_{z,2}}(\omega_0)z),  &\,|z| < \frac{a}{2}, \\ 
  &\cos \left({k_{z,2}}(\omega_0)\frac{a}2\right)\ee^{\ii k_{z,1}(\omega_0)(|z| - a/2)}, &|z| > \frac{a}{2} \\ 
 \end{aligned}\right.
\end{equation}
describes the dependence of the envelope of the reflected pulse on the $z$~coordinate.
From Eqs.~\eqref{eq:20} and~\eqref{eq:21}, it is easy to see that the envelope of the $z$-component of the magnetic field of the reflected pulse possesses a spatiotemporal optical vortex at each $z$~value.
Moreover, within the linear approximation of Eq.~\eqref{eq:15}, the components $E_y$ and $H_x$, which for each fixed value of $z$ are proportional to the envelope of the $H_z$ component, will also contain STOVs.

The presented analysis demonstrates the generation of a spatiotemporal optical vortex in the components of the envelope of the reflected pulse propagating in the waveguide using a diffractive structure with the TF of Eqs.~\eqref{eq:13},~\eqref{eq:18}, and~\eqref{eq:19} in the case of a Gaussian envelope of the incident pulse~\eqref{eq:16} [and, accordingly, a Gaussian spectrum~\eqref{eq:17}].
It is worth noting that the generation of a vortex can also be demonstrated for a more general class of incident pulses having the spectrum with the following $(k_y,\tilde\omega )$-dependence:
\begin{equation}
\label{eq:22}
	G(k_y, \tilde\omega ) = {\hat G}(\sigma_y^2 k_y^2 + \sigma_t^2 { \tilde\omega}^2).
\end{equation}
Indeed, for the sake of simplicity, let us consider the envelope of the reflected pulse $P_{r,H_z}(0, y_r, z, t)$ at $z=0$.
In this case, at the TF of Eqs.~\eqref{eq:13},~\eqref{eq:18}, and~\eqref{eq:19}, we obtain:
\begin{equation}
\label{eq:23}
\begin{aligned}
P_{r,H_z}(0,y_r,0,t) = \frac{c_y}{\sigma_y}
\iint &{\hat G}(\sigma_y^2 k_y^2 + \sigma_t^2 {\tilde\omega}^2)
\cdot
(\sigma_y k_y \pm \ii \sigma_t \tilde\omega)
\\&\cdot\ee^{\ii k_y y_r - \ii \tilde\omega t} 
\dd{k_y}\dd{\tilde\omega}.
\end{aligned}
\end{equation}
By transforming Eq.~\eqref{eq:23} to generalized polar coordinates in the frequency domain $\sigma_y k_y = \rho \cos \varphi$, $\sigma_t \tilde\omega = \rho \sin \varphi$ and also using generalized polar coordinates in the spatiotemporal domain $y/{\sigma_y} = s\cos \psi$, $t/{\sigma_t} = s\sin \psi$, we arrive at
\begin{equation}
\label{eq:24}
P_{r, H_z}(s, \psi) = 2\ii\pi \frac{c_y}{\sigma_y}\ee^{\mp \ii\psi}
\int_0^\infty {\hat G}(\rho^2){\rm J}_1(s \rho ){\rho^2} \dd\rho,
\end{equation}
where ${\rm J}_1(\rho)$ is the Bessel function of the first kind of order one. 
The presence of the term $\ee^{\mp \ii\psi}$ and the fact that ${\rm J}_1(0) = 0$ mean that the envelope of the reflected pulse~\eqref{eq:24} contains an STOV in the coordinates $(s, \psi)$, which, according to Eqs.~\eqref{eq:20} and~\eqref{eq:21}, exists at each $z$ value.

It is instructional to consider the evolution of the generated STOV during the further propagation of the reflected pulse.
In the general case, the propagation of the envelope of the reflected pulse is described by Eq.~\eqref{eq:11}.
This equation, however, does not provide a closed-form expression for the STOV pulse shape at a fixed value of $x_r$ (or at fixed $t$).
Thus, let us consider approximate models for the pulse propagation that describe the STOV evolution.
In the simplest case, we expand the wave vector component $k_x(\tilde\omega, k_y)$ up to the linear terms:
\begin{equation}
\label{eq:25}
\begin{aligned}
k_x(\tilde\omega, k_y) &= \sqrt {{{\left( {\frac{\omega_0+\tilde\omega}{\cc}{n_{\rm eff}}(\omega_0+\tilde\omega)} \right)}^2} - k_y^2} 
\\&\approx \frac{\omega_0}{\cc}{n_{\rm eff}}(\omega_0) + \frac{\tilde\omega}{v},
\end{aligned}
\end{equation}
where $v = \left( \partial k_x / \partial \tilde\omega \right)^{-1} = \cc{\left[ {{n_{\rm eff}}(\omega_0) + \omega_0{n'_{\rm eff}}(\omega_0)} \right]^{-1}}$ is the group velocity. By substituting Eq.~\eqref{eq:25} into Eq.~\eqref{eq:11}, we will obtain a pulse propagating with the group velocity $v$ with a conserved envelope shape:
\begin{equation}
\label{eq:26}
\begin{aligned}
{{\bf P}_r}\left(y_r, z,t + \frac{x_r}{v}\right) 
= \iint &G(k_y, \tilde\omega ){\bf\Psi}\left( {\omega_0+\tilde\omega, k_y, z} \right)H(k_y, \tilde\omega )
\\&\cdot\ee^{\ii k_y y_r}\ee^{ -\ii\tilde\omega (t + x_r/v)} \dd{k_y}\dd{\tilde\omega }.
\end{aligned}
\end{equation}
Therefore, in the linear approximation of Eq.~\eqref{eq:25}, the generated STOV is obviously conserved upon propagation.

This approximate model, however, neglects the spatial and temporal broadening of the pulse during propagation. 
Therefore, let us consider a more accurate quadratic approximation
\begin{equation}
\label{eq:27}
	k_x(\tilde\omega, k_y) \approx \frac{\omega_0}{\cc}{n_{\rm eff}}(\omega_0) + \frac{\tilde\omega}{v} + \alpha {\tilde\omega ^2} + \beta k_y^2,
\end{equation}
where $2\alpha = \frac{{\partial ^2 k_x}}{{\partial {\tilde\omega}^2}}(0,0) = \left[ {2 n'_{\rm eff}(\omega_0) + \omega_0 n''_{\rm eff}(\omega_0)} \right]/(2\cc)$ is the group velocity dispersion describing the temporal broadening and the quantity $2\beta = \frac{\partial^2{k_x}}{\partial k_y^2}(0,0) = - {\left[ {\omega_0{n_{\rm eff}}(\omega_0)/\cc} \right]^{-1}}$ describes the spatial broadening.
By substituting Eq.~\eqref{eq:27} into Eq.~\eqref{eq:11}, we will obtain that in the quadratic approximation, the propagation of the envelope of the reflected pulse $P_{r, H_z}$ generated at the TF of Eqs.~\eqref{eq:13},~\eqref{eq:18}, and~\eqref{eq:19} will be described by the expression
\begin{equation}
\label{eq:28}
\begin{aligned}
	P_{r, H_z}(x_r, y_r, 0, t) = \iint &G(k_y, \tilde\omega )c_y 
	\left(k_y \pm \ii\frac{\sigma_t}{\sigma_y} \tilde\omega\right)
	\ee^{ -\ii{x_r}\left( {\alpha {{\tilde\omega }^2} + \beta k_y^2} \right)}
	\\&\cdot\ee^{-\ii\tilde\omega (t + {x_r}/v) + \ii k_y y_r}\dd{k_y}\dd\tilde\omega.
	\end{aligned}
\end{equation}
In the case of a Gaussian incident pulse [Eq.~\eqref{eq:17}], the latter integral can be easily calculated as
\begin{equation}
\label{eq:29}
\begin{aligned}
P_{r,H_z}(x_r, y_r, 0, t) &= 2\ii\frac{c_y}{\sigma_y}
\left( \frac{y_r}{\sigma _y\gamma_y} \mp \ii\frac{t + x_r/v}{\sigma_t\gamma_t}\right)
\\&\cdot\frac{1}{\sqrt{\gamma_y \gamma_t}}
\exp \left\{ -\frac{y_r^2}{\sigma_y^2 \gamma_y} - \frac{(t + x_r/v)^2}{\sigma_t^2 \gamma_t} \right\},
\end{aligned}
\end{equation}
where ${\gamma_y} = 1 + 4\ii{x_r}\beta /\sigma_y^2$ and ${\gamma_t} = 1 + 4\ii{x_r}\alpha /\sigma_t^2$.
Note that at $x_r=0$, Eq.~\eqref{eq:29} coincides with the previously obtained Eq.~\eqref{eq:20} describing the generation of an STOV at $x_r=0$.
From Eq.~\eqref{eq:29}, one can easily see that the reflected pulse contains a phase singularity at $y_r=0$ and $t = -x_r/v$.
Thus, one can conclude that in the quadratic approximation, the phase singularity generated at $x_r=0$ then propagates with the velocity $v$ in the negative direction of the $x_r$ axis.

\section{Integrated metal-dielectric structure for the generation of an STOV}
Having theoretically described the conditions for generating a spatiotemporal optical vortex in a slab waveguide, let us now discuss the design of an integrated structure satisfying these conditions.
The first condition is the presence of a reflection zero at $(k_y, \tilde\omega) = (0,0)$ (let us remind that $k_y$ is the transverse wavevector component in the coordinate system associated with the incident optical signal and $\tilde\omega = \omega - \omega_0$).
In other words, this condition means that the used photonic structure must possess a reflection zero at certain ``central'' angular frequency $\omega_0$ and angle of incidence $\theta$ [see Fig.~\ref{fig:1} and Eq.~\eqref{eq:10}].
The second condition, Eq.~\eqref{eq:18}, requires that the difference between the arguments of the complex coefficients ${c_t}$ and ${c_y}$ at the linear terms of the transfer function of the structure [see Eq.~\eqref{eq:13}] equals $\pm\pi/2$.

As an example, let us consider a 100 nm thick gallium phosphide (GaP) waveguide layer sandwiched between symmetric silicon dioxide (${\rm SiO}_2$) claddings (see Fig.~\ref{fig:1}).
At the angular frequency $\omega_0 = 2.9899 \cdot 10^{15}\sinv$ corresponding to the free-space wavelength of 630 nm, this waveguide is single-mode and supports fundamental TE- and TM-polarized guided modes with effective refractive indices $n_{\rm eff} = 2.773$ and $n_{\rm eff, TM} = 2.137$, respectively.
For an incident TE-polarized mode, there is no out-of-plane scattering (scattering to the claddings) for angles of incidence greater than $31.7^\circ$, and no excitation of reflected and transmitted TM-polarized modes at angles of incidence greater than $50.4^\circ$.
Moreover, if the whole integrated structure possesses a horizontal symmetry plane (as in the case considered in the present work), scattered TM-polarized modes will not be excited regardless of the angle of incidence~\cite{32}.

As an integrated nanophotonic element intended for the generation of an STOV in the waveguide layer, we propose to use a ``three-strip'' metal-dielectric structure consisting of three gold (Au) strips ``buried'' in the waveguide (Fig.~\ref{fig:1}).
The thickness of the strips equals the thickness of the core waveguide layer.
Let us explain the choice of this geometry of the structure.
For the free-space radiation, one of the simplest structures, which enable obtaining a reflection zero, is a three-layer metal-dielectric-metal (MDM) resonator consisting of two metal layers separated by a dielectric layer.
In~\cite{33,34}, it was shown that for an MDM structure, at fixed parameters of the incident plane wave (angle of incidence, frequency, and polarization) and at fixed thickness of the lower metal layer, a reflection zero can always be obtained by choosing the thicknesses of the upper metal layer and the dielectric layer.
The appearance of the reflection zero has a resonant nature and is caused by the excitation of a mode localized between the metal layers.
An integrated counterpart of the MDM structure consisting of two metal strips in the waveguide layer was investigated in the recent work~\cite{35} by some of the present authors, in which the possibility of obtaining zero reflectance in the ``scattering-free'' regime was demonstrated.

Since for the STOV generation, in addition to the zero reflection, the $\pm\pi/2$ phase difference between the linear terms of the transfer function of the structure (represented by the reflection coefficient) has to be provided, in the present work we consider a more complex version of the integrated MDM structure containing an additional metal strip.
The parameters of the proposed three-strip structure are the widths of the metal strips and of the dielectric segments of the waveguide separating them $w_i,\;i = 1\ldots5$ (Fig.~\ref{fig:1}).
The widths $w_i$ were optimized using a genetic algorithm to satisfy the STOV generation conditions discussed above.
The optimization involved multiple solution of the ``direct'' diffraction problem (simulation of the diffraction of the TE-polarized incident mode on the structure), which was carried out using an in-house implementation of the aperiodic Fourier modal method adapted to the solution of the problems of integrated optics~\cite{36}.
As a result, the following structure satisfying the STOV generation conditions at $\theta = 65^\circ$ was obtained:
\begin{equation}
\label{eq:30}
	(w_1, w_2, w_3, w_4, w_5) = (25.4,52.1,60,170,120)\nm.
\end{equation}

\begin{figure*}[th]
\centering\includegraphics{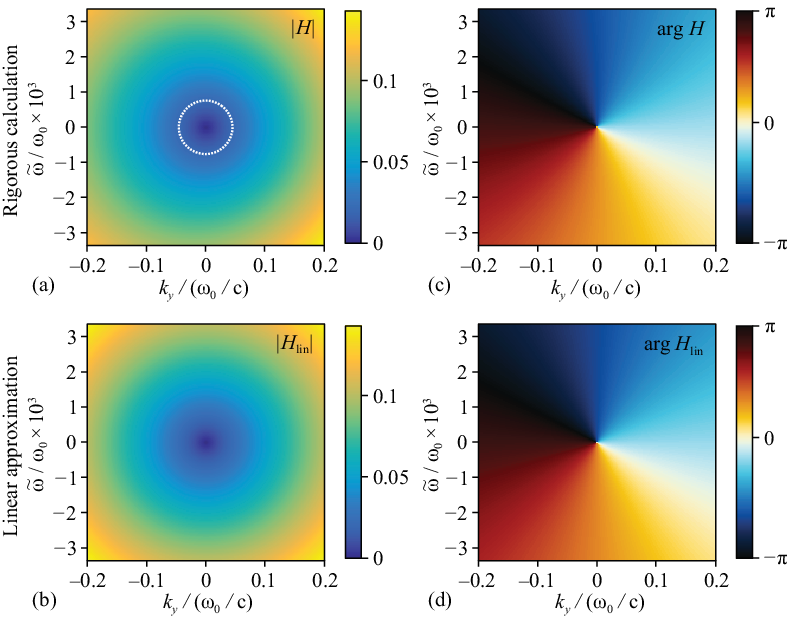}
\caption{\label{fig:2}Absolute values~(a),~(b) and arguments~(c),~(d) of the numerically calculated TF of the investigated structure~(a),~(c) and its linear approximation of Eq.~\eqref{eq:13}~(b),~(d). Dotted white contour in~(a) shows the spectrum of the incident signal at the $1 /\ee^2$ level.
}
\end{figure*}

Figure~\ref{fig:2} shows the absolute value [Fig.~\ref{fig:2}(a)] and argument [Fig.~\ref{fig:2}(c)] of the numerically calculated TF of this structure ${H_{\rm calc}}$ and of the ``model'' transfer function of Eq.~\eqref{eq:13} [Figs.~\ref{fig:2}(b) and~\ref{fig:2}(d)].
The latter function was plotted for the values $c_y = 0.046\exp (  -0.4518\ii )\um$ and $c_t = 9.36\exp ( -2.0247\ii )\fs$.
These values were obtained by fitting to the numerically calculated data and provide $\arg (c_t/c_y) \approx -\pi/2$.
The normalized root-mean-square deviation between the complex-valued TF and its linear approximation in the angular and spatial frequency ranges shown in Fig.~\ref{fig:2} amounts to only 5.26\%.
 
\section{Generation of an STOV using the designed metal-dielectric structure}
\begin{figure*}[tbh]
\centering\includegraphics{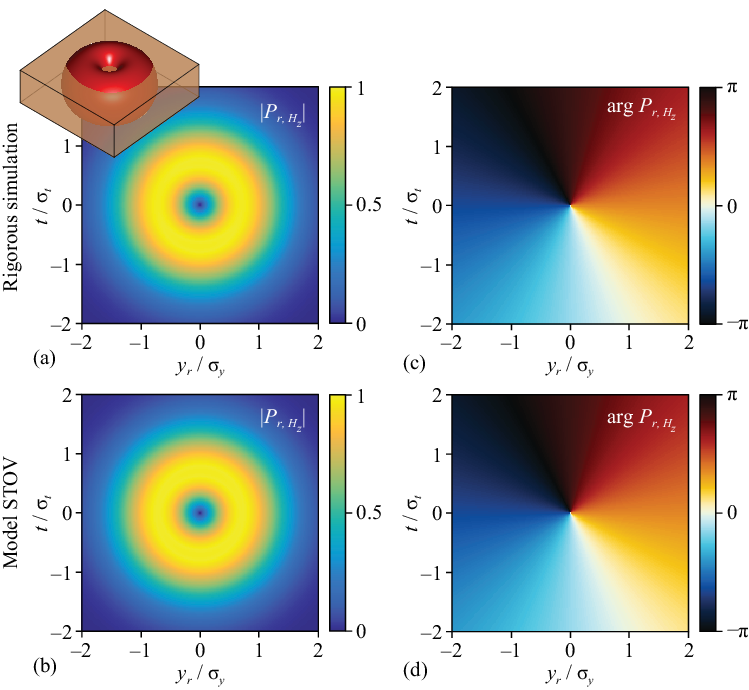}
\caption{\label{fig:3}Normalized amplitudes (absolute values)~(a),~(b) and phases (arguments)~(c),~(d) of the numerically calculated envelope of the $H_z$ component of the reflected optical pulse~(a),~(c) and of the model function of Eq.~\eqref{eq:20}~(b),~(d) at $z = 0$. The inset to~(a) shows the envelope of the reflected pulse in the coordinates $\left(y_r / \sigma_y, t /\sigma_t,z \right)$ at the $10^{-3}$ level. The rectangular cuboid shown in the inset depicts the waveguide.
}
\end{figure*}

In order to demonstrate the generation of a spatiotemporal optical vortex using the designed structure, let us consider an incident Gaussian beam [see Eqs.~\eqref{eq:16} and~\eqref{eq:17}] with $\sigma_y = 14\um$ and $\sigma_t = 2.8\ps$ providing the fulfillment of Eq.~\eqref{eq:19} [the $1 /\ee^2$ level of the spectrum~\eqref{eq:17} of the incident signal is shown with a white dotted line in Fig.~\ref{fig:2}(a)].
Figures~\ref{fig:3}(a) and~\ref{fig:3}(c) show the absolute value and phase of the envelope of the $H_z$ component of the reflected pulse at the central plane of the waveguide core layer (at $z = 0$) calculated numerically using the rigorous Eq.~\eqref{eq:11}.
From these figures, it is evident that the reflected optical signal indeed contains a spatiotemporal optical vortex.
The inset to Fig.~\ref{fig:3}(a) shows the ``three-dimensional'' envelope (isosurface at the $10^{-3}$ level) of the reflected pulse in the coordinates $(y_r / \sigma_y, t /\sigma_t,z)$.
For the sake of comparison, Figs.~\ref{fig:3}(b) and~\ref{fig:3}(d) show the ``model'' envelope $P_{r, H_z}$ of Eq.~\eqref{eq:20}.
The numerically calculated and model envelopes are in a good agreement: the normalized root-mean-square deviation between them amounts to only~0.37\%.

\begin{figure*}[tp]
\centering\includegraphics{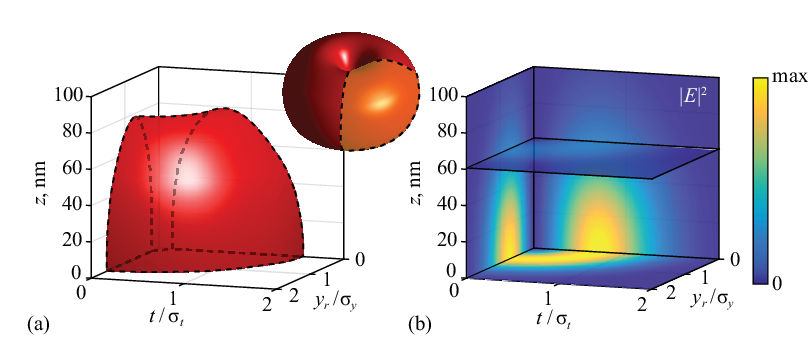}
\caption{\label{fig:4}Electromagnetic field intensity of the reflected pulse at the level of~$0.1$ of its maximum value~(a) and cross-sections of the normalized intensity distribution~(b) in the first octant of the coordinate system $( y_r / \sigma_y, t /\sigma_t,z )$. The inset to~(a) shows the ``full'' isosurface of the 3D intensity distribution.
}
\end{figure*}

As it was shown at the end of Section~\ref{sec:2}, within the linear approximation of Eq.~\eqref{eq:15}, all the components of the electromagnetic field of the reflected pulse at each fixed value of $z$ are proportional to the envelope of the $H_z$ component, and therefore also contain an STOV.
In order to illustrate this fact, Fig.~\ref{fig:4} shows the calculated isosurface of the intensity of the electric field of the reflected pulse $I(\tilde y,\tilde t,z) = \left| {\bf E}_r(\tilde y, \tilde t, z) \right|^2$, where ${{\bf E}_r}(\tilde y,\tilde t,z)$ is the electric field vector of the reflected pulse at $x_r=0$ [Fig.~\ref{fig:4}(a)], and the cross-sections of the intensity distribution in the planes $z=0$, $z=60\nm$, $\tilde y = y_r / \sigma_y = 0$, and $\tilde t = t/\sigma_t = 0$ [Fig.~\ref{fig:4}(b)].
For illustrative purposes, the isosurface and the cross-sections are shown in the first octant of the coordinate system $(\tilde y, \tilde t, z)$.
Let us note that although in the approximation of Eq.~\eqref{eq:15}, the only nonzero electric field component is the $E_y$ component, the intensity was calculated using the rigorous Eq.~\eqref{eq:11} taking into account the $E_x$ component.
From Fig.~\ref{fig:4}, it is evident that the isosurface of the intensity distribution has a doughnut-like shape, and the intensity distribution vanishes at $y_r=0$ and $t=0$ at all $z$ values, which confirms the generation of an STOV in the electric field components of the reflected pulse.

\begin{figure*}[th]
\hspace{-7em}
\includegraphics{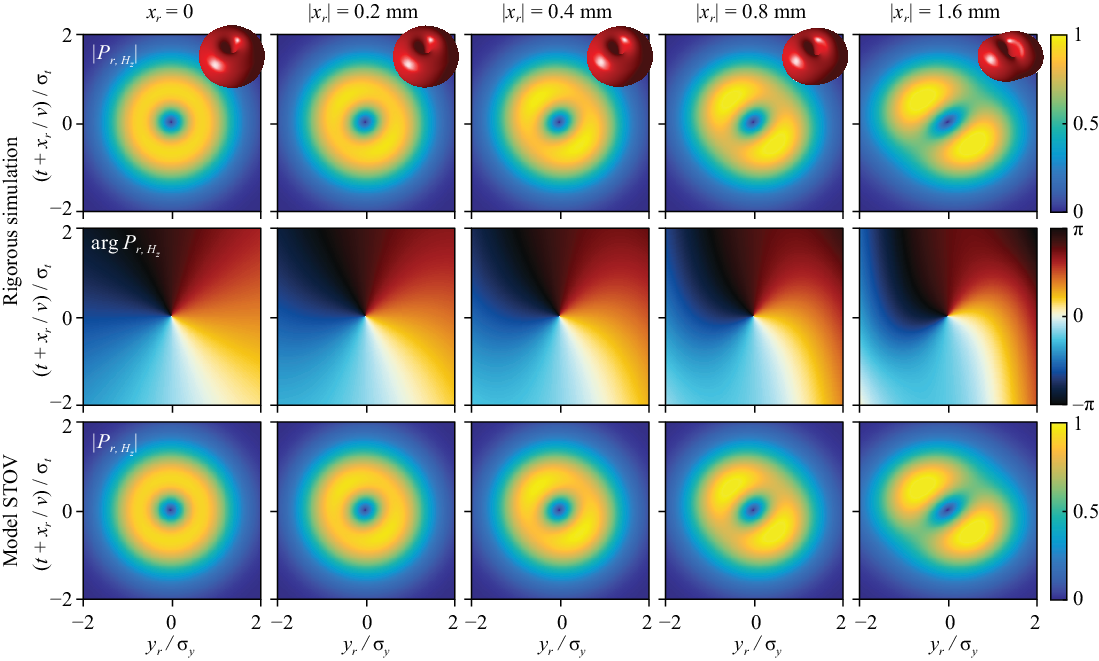}
\caption{\label{fig:5}Normalized absolute values (upper row) and phases (middle row) of the envelope $P_{r, H_z}$ of the reflected pulse at different distances from the structure calculated using the rigorous Eq.~\eqref{eq:11} and the absolute values of the ``model'' envelope of Eq.~\eqref{eq:29} (bottom row). The insets to the upper row show the envelopes of the reflected pulse in the coordinates $(y_r/ \sigma_y, t/\sigma_t, z)$ at different distances $|x_r|$.
}
\end{figure*}

In addition, the propagation of the reflected pulse containing an STOV was investigated.
The upper two rows of Fig.~\ref{fig:5} show the evolution of the envelope $P_{r, H_z}$ of the reflected pulse upon propagation.
The calculation of the envelopes in this case was carried out using the rigorous Eq.~\eqref{eq:11} at different propagation distances $|x_r|$.
The envelopes are shown in normalized coordinates $\tilde y = y_r/{\sigma_y}$ and $\tilde t = (t + x_r/v)/{\sigma_t}$, where $v = 0.2378\cc$ is the group velocity.
One can see that at $|x_r| \leq 0.2\mm$, the shape of the envelope changes only very slightly, which corresponds to the linear approximation of Eqs.~\eqref{eq:25} and~\eqref{eq:26}.
With an increase in the propagation distance, the envelope becomes ``distorted'', however, the location of the center of the STOV is almost perfectly described by the expression $t = -x_r/v$.

The bottom row of Fig.~\ref{fig:5} shows the absolute values of the envelopes of the reflected pulse calculated using the analytical expression~\eqref{eq:29} (the corresponding phases are almost visually identical to the rigorously calculated ones depicted in the middle row and are therefore not shown).
Let us remind that Eq.~\eqref{eq:29} was obtained in the quadratic approximation~\eqref{eq:27} for the case of an ``ideal'' TF of Eqs.~\eqref{eq:13},~\eqref{eq:18}, and~\eqref{eq:19}.
Fig.~\ref{fig:4} demonstrates a good correspondence between the rigorous calculation results and the ``model'' Eq.~\eqref{eq:29}.
In particular, at $|x_r| = 1.6\mm$, the normalized root-mean-square deviation between the envelopes calculated using Eqs.~\eqref{eq:11} and~\eqref{eq:29} amounts to only 0.65\%.
Thus, Eq.~\eqref{eq:29} can be considered as a quite accurate ``model'' expression for describing the STOV propagation.

\section{Conclusion}
We investigated theoretically and demonstrated numerically the generation of an STOV beam in a dielectric slab waveguide.
In the proposed integrated geometry, the generation of an STOV beam is performed upon reflection of a spatiotemporal optical pulse with a Gaussian envelope propagating in the waveguide from a metal-dielectric structure consisting of several metal strips ``buried'' in the waveguide core layer.
The interaction of the incident pulse with the metal-dielectric structure is described with the use of a ``vectorial'' spatiotemporal transfer function obtained in the framework of the theory of linear systems.
We demonstrated that if the TF of the structure has a linear form with respect to the spatial and angular frequencies, and the complex coefficients at the linear terms have a $\pi/2$ argument difference, then the integrated structure generates a reflected pulse containing an STOV in all nonzero components of the electromagnetic field.
The rigorous numerical simulation results of the designed three-strip metal-dielectric structure fully confirmed the obtained theoretical results and demonstrated the generation of an STOV in the slab waveguide.

In our opinion, the presented results will facilitate the research and application of STOV beams in the on-chip geometry.
In particular, we believe that the obtained results can be extended to other platforms of integrated photonics, in particular, the platform of Bloch surface waves supported by interfaces of photonic crystals.

\section*{Funding}
Russian Science Foundation (19-19-00514); 
State assignment to the FSRC ``Crystallography and Photonics'' RAS.

\section*{Acknowledgments}
This work was funded by the Russian Science Foundation (theoretical and numerical investigation of on-chip STOV generation using integrated metal-dielectric structures) and performed within the State assignment to the FSRC ``Crystallography and Photonics'' RAS (numerical simulation of the STOV evolution during propagation).

\section*{Disclosures}
 The authors declare no conflicts of interest.

\section*{Data Availability}
Data underlying the results presented in this paper are not publicly available at this time but may be obtained from the authors upon reasonable request.


\end{document}